\documentclass[preprint]{jpsj2}

\usepackage{latexsym}
\usepackage{graphicx}

\def\gs2{0.30} 
\def\g3{1.00} 

\title{Statistical Mechanics of 
Linear and Nonlinear \\
Time-Domain Ensemble Learning}

\author{
\textsc{Seiji MIYOSHI}$^{1}$
\thanks{E-mail address: miyoshi@kobe-kosen.ac.jp}
and \textsc{Masato OKADA}$^{2}$$^{3}$$^{4}$}

\inst{
$^{1}$Department of Electronic Engineering, 
Kobe City College of Technology, \\
8--3 Gakuenhigashimachi, Nishi-ku, Kobe-shi, 651--2194 \\
$^{2}$Division of Transdisciplinary Sciences, 
Graduate School of Frontier Sciences, \\
The University of Tokyo, 
5--1--5 Kashiwanoha, Kashiwa-shi, Chiba, 277--8561 \\
$^{3}$RIKEN Brain Science Institute, 
2--1 Hirosawa, Wako-shi, Saitama, 351--0198
}

\abst{
Conventional ensemble learning combines students
in the space domain. 
In this paper, however, we combine 
students in the time domain
and call it time-domain ensemble learning.
We analyze, compare, and discuss
the generalization performances regarding 
time-domain ensemble learning of 
both a linear model and a nonlinear model.
Analyzing 
in the framework of online learning 
using a statistical mechanical method,
we show the qualitatively different behaviors
between the two models.
In a linear model,
the dynamical behaviors of the generalization error
are monotonic.
We analytically show that time-domain
ensemble learning is twice as effective as 
conventional ensemble learning.
Furthermore, the generalization error of 
a nonlinear model
features nonmonotonic dynamical behaviors
when the learning rate is small.
We numerically show that the generalization performance 
can be improved remarkably by
using this phenomenon and the divergence of 
students in the time domain.
}

\kword{
ensemble learning, 
online learning, 
generalization error,
statistical mechanics
}

\begin{document}


\maketitle

\section{Introduction}
Learning can be roughly classified into batch learning and
online learning \cite{Saad}.
In batch learning, given examples are used more than once.
In this paradigm, a student will give the correct answers
after training if that student has an adequate degree of freedom.
However, it is necessary to have plenty of time and 
a large memory for storing many examples.
On the contrary, in online learning examples 
used once are then discarded.
In this case, a student cannot give correct answers 
for all the examples used in training.
There are, however, merits.
For example,
a large memory for storing many examples is not necessary,
and it is possible to follow a time-variant teacher. 

Recently, we analyzed the generalization performance
of some models
in a framework of online learning
using a statistical mechanical method
\cite{Hara,PRE,JPSJ2006a,JPSJ2006b,JPSJ2006c}.
Ensemble learning means 
to combine many rules or learning machines
(called students in this paper) that perform poorly;
this has recently attracted the attention 
of many researchers
\cite{Hara,PRE,Abe,www.boosting.org,Krogh,Urbanczik}.
The diversity or variety of students is essential
in ensemble learning. 
We showed that
the three well-known rules, 
Hebbian learning, perceptron learning, and
AdaTron learning
have different characteristics in their 
affinities for ensemble learning, that is in 
``maintaining diversity among students" \cite{PRE}．
In that process \cite{IBIS2004,NC200503}, 
it was subsidiarily proven 
that in an unlearnable case \cite{Inoue,Inoue2},
the student vector 
does not converge in one direction 
but continues moving.
Therefore, 
we also analyzed the generalization performance
of a student supervised by a moving teacher that goes around
a true teacher \cite{JPSJ2006a},
proving 
that
the generalization error of a student
can be smaller than
a moving teacher's,
even if the student only uses examples 
from the moving teacher.
In an actual human society, a teacher
observed by a student
does not always present the correct answer.
In fact, many cases the teacher is learning
and continues to change.
Therefore, analyzing such a model
is interesting in terms of considering the 
analogies between statistical learning theories 
and real human society.

In conventional ensemble learning,
generalization performance is improved by combining
students who have diversities.
However, students do not always
converge in one direction but may continue moving 
in an unlearnable model.
Therefore, generalization performance in such a model 
must be improved by combining students themselves
at different times, even if there is only one student.
Conventional ensemble learning combines students
in the space domain. 
In contrast, we introduce a method of combining 
the students in the time domain,
which we call 
\lq\lq time-domain ensemble learning\rq\rq
\cite{JPSJ2006c}.

Some studies \cite{Dekel,Freund,Cesa2004,Cesa2005} 
have treated the combining of students in the time domain.
We particularly pay attention to dynamical behaviors 
of the generalization performance 
of the time-domain ensemble learning
and 
theoretically analyze it by applying a statistical 
mechanical method.
We analytically or numerically obtain, compare, and discuss
the order parameters and the generalization 
errors of two models: 
a linear model in which both teacher and student
are linear perceptrons \cite{Hara} with noise
and a nonlinear model in which both teacher and student
are nonlinear perceptrons.
The results 
show that the two models have 
the qualitatively different behaviors.
We analytically demonstrate that 
time-domain ensemble learning of a linear model
is twice as effective as 
conventional ensemble learning.
Furthermore, 
we numerically show that
the generalization performance 
of a nonlinear model
can be ramarkably improved by
using nonmonotonic dynamical behaviors and
the divergence of students in the time domain.

\section{Model}
In this paper we consider a teacher and a student.
They are perceptrons with the connection weights
$\mbox{\boldmath $B$}$ and $\mbox{\boldmath $J$}^m$, 
respectively,
where $m$ denotes the time step.
For simplicity, the connection weights of the teacher
and the student
are simply called the teacher and the student.
Teacher $\mbox{\boldmath $B$}=\left(B_1,\ldots,B_N\right)$,
student $\mbox{\boldmath $J$}^m=\left(J^m_1,\ldots,J^m_N\right)$,
and input
$\mbox{\boldmath $x$}^m=\left(x^m_1,\ldots,x^m_N\right)$
are $N$-dimensional vectors.
Each component $B_i$ of $\mbox{\boldmath $B$}$
is independently drawn from ${\cal N}(0,1)$ and fixed,
where ${\cal N}(0,1)$ denotes a Gaussian distribution with
a mean of zero and a variance of unity.
Each component $J_i^0$
of the initial value $\mbox{\boldmath $J$}^0$
of $\mbox{\boldmath $J$}^m$
is independently drawn from ${\cal N}(0,1)$.
The direction cosine between 
$\mbox{\boldmath $J$}^m$ and 
$\mbox{\boldmath $B$}$ is $R^m$
and that between
$\mbox{\boldmath $J$}^m$ and
$\mbox{\boldmath $J$}^{m'}$ is $q^{m,m'}$.
Each component $x^m_i$ of $\mbox{\boldmath $x$}^m$
is drawn from ${\cal N}(0,1/N)$ independently.

\begin{figure}[htbp]
\vspace{3mm}
\begin{center}
\includegraphics[width=0.3\linewidth,keepaspectratio]{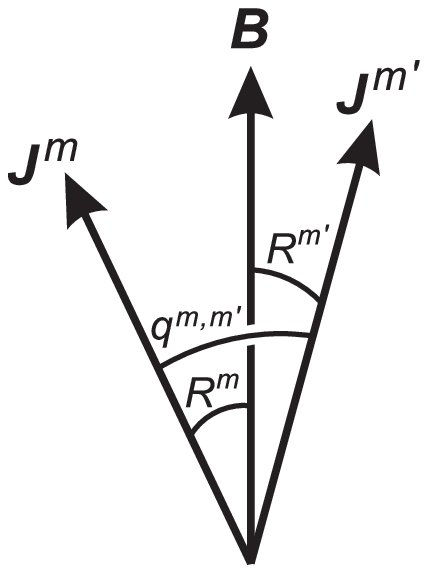}
\caption{
Teacher $\mbox{\boldmath $B$}$ and
students $\mbox{\boldmath $J$}^m$
and $\mbox{\boldmath $J$}^{m'}$.
$R^m, R^{m'}$, and $q^{m,m'}$ are direction cosines.
}
\label{fig:BJJ}
\end{center}
\end{figure}

Figure \ref{fig:BJJ} illustrates
the relationship among teacher
$\mbox{\boldmath $B$}$,
students $\mbox{\boldmath $J$}^m$ and 
$\mbox{\boldmath $J$}^{m'}$
and the direction cosines
$R^m, R^{m'}$, and $q^{m,m'}$.

In this paper, we also deal with 
the thermodynamic limit $N\rightarrow \infty$. 
Therefore,
$\|\mbox{\boldmath $B$}\|=\sqrt{N},\ 
\|\mbox{\boldmath $J$}^0\|=\sqrt{N},$ and
$\|\mbox{\boldmath $x$}^m\|=1$.
Generally, since the norm $\|\mbox{\boldmath $J$}^m\|$
of the student
changes as the time step proceeds,
the ratios $l^m$ of the norm to $\sqrt{N}$
are introduced and are called the length of 
the student. That is,
$\|\mbox{\boldmath $J$}^m\|=l^m\sqrt{N}$.

\begin{description}
\item[Linear Case]
Outputs of the teacher and the student
are 
$o_B^m=v^m+n_B^m$ and $o_J^m=u^ml^m+n_J^m$, respectively.
Here,
$
v^m = 
 \mbox{\boldmath $B$}\cdot \mbox{\boldmath $x$}^m, 
u^m l^m = 
 \mbox{\boldmath $J$}^m\cdot \mbox{\boldmath $x$}^m, 
n_B^m \sim {\cal N}\left(0,\sigma_B^2\right),
n_J^m \sim {\cal N}\left(0,\sigma_J^2\right)
$,
where
${\cal N}(0,\sigma^2)$ denotes a Gaussian distribution with
a mean of zero and a variance of $\sigma^2$.
That is,
the outputs of the teacher and the student
include independent Gaussian noises with variances of 
$\sigma_{B}^2$ and $\sigma_{J}^2$, respectively.
Then, $v^m$ and $u^m$
obey Gaussian distributions with a mean of zero,
a variance of unity, and a covariance of $R^m$.
Let us define the error $\epsilon^m_S$ between 
the teacher $\mbox{\boldmath $B$}$
and the student $\mbox{\boldmath $J$}^m$ alone
by the squared error of their outputs:
\begin{equation}
\epsilon^m_S \equiv 
\frac{1}{2}\left(o_B^m-o_J^m\right)^2.
\label{eqn:eS}
\end{equation}

Student $\mbox{\boldmath $J$}^m$
adopts the gradient method as a learning rule
and uses
input $\mbox{\boldmath $x$}$
and an output
of teacher
$\mbox{\boldmath $B$}$
for updates.
That is,
\begin{eqnarray}
\mbox{\boldmath $J$}^{m+1}
&=& \mbox{\boldmath $J$}^{m} 
   -\eta \frac{\partial \epsilon^m_S}{\partial \mbox{\boldmath $J$}^{m}}\\
&=& \mbox{\boldmath $J$}^{m} 
   +\eta \left( v^m+n_B^m-u^ml^m-n_J^m\right)
   \mbox{\boldmath $x$}^{m}, \label{eqn:Jupdate-l}
\end{eqnarray}
where $\eta$ denotes the learning rate of the student
and is a constant positive number.
The part $\eta \left( v^m+n_B^m-u^ml^m-n_J^m\right)$
has been determined by the learning rule.
Generalizing this part, 
we denote it with $f^m$.

\item[Nonlinear Case]
The teacher and the student are 
a nonmonotonic perceptron and 
a simple perceptron, respectively;
their outputs are 
$
o_B^m=\mbox{sgn}\left(\left(v^m-a\right)v^m\left(v^m+a\right)\right),\ 
o_J^m=\mbox{sgn}\left(u^ml^m\right)
$.
Here, 
$v^m=\mbox{\boldmath $B$}\cdot\mbox{\boldmath $x$}^m$ and 
$u^ml^m=\mbox{\boldmath $J$}^m \cdot\mbox{\boldmath $x$}^m$,
$v^m$ and $u^m$
obey Gaussian distributions with a mean of zero,
a variance of unity, and a covariance of $R^m$, and
$\mbox{sgn}(\cdot)$ denotes a sign function.
Student $\mbox{\boldmath $J$}^m$
adopts the perceptron learning as a learning rule for updates.
That is,
\begin{eqnarray}
\mbox{\boldmath $J$}^{m+1}
&=& \mbox{\boldmath $J$}^{m} +
   \eta \Theta\left(-o_B^m o_J^m\right)o_B^m
   \mbox{\boldmath $x$}^{m}, \label{eqn:Jupdate-n}
\end{eqnarray}
where $\eta$ denotes the learning rate of the student
and is a constant positive number,
$\Theta(\cdot)$ denotes a step function.
The part 
$\eta \Theta\left(-o_B^m o_J^m\right)o_B^m$
has been determined by the learning rule.
Generalizing this part, we denote it with $f^m$.


\end{description}

\section{Theory}
\subsection{Generalization error}
Conventionally, ensemble learning means 
to improve performance
by combining many students that perform poorly.
We, however, use 
just one student and combine copies of it
(hereafter called \lq\lq  brothers\rq\rq)
at different time steps in this paper.
Conventional ensemble learning combines students
in the space domain, whereas, we 
do so in the time domain.
In this paper $K$ brothers
$\mbox{\boldmath $J$}^{m_1}, \mbox{\boldmath $J$}^{m_2},\ldots,
\mbox{\boldmath $J$}^{m_K}$
are combined.
Here, $m_1 \leq m_2 \leq \ldots \leq m_K$.
One goal of statistical learning theory
is to theoretically obtain generalization errors.
Since a generalization error is the mean of errors 
over the distribution of the new input $\mbox{\boldmath $x$}$
,
the generalization errors $\epsilon_g$ of the ensemble
in linear and nonlinear cases are
calculated as follows:

\begin{description}
\item[Linear Case]
We use the squared error $\epsilon$
for new input $\mbox{\boldmath $x$}$.
Here, it is assumed that
the Gaussian noises 
are independently added to 
the teacher and each brother of the ensemble.
The weight of each brother $\mbox{\boldmath $J$}^{m_k}$ 
of the ensemble satisfies $C_k \ge 0$.
That is, the error of the ensemble is
\begin{equation}
\epsilon=\frac{1}{2}
\left(\mbox{\boldmath $B$}\cdot\mbox{\boldmath $x$}+n_B
-\sum_{k=1}^K C_k 
\left(\mbox{\boldmath $J$}^{m_k}\cdot\mbox{\boldmath $x$}+n_k \right)
\right)^2,
\label{eqn:e}
\end{equation}
where
$n_B \sim {\cal N}\left(0,\sigma_B^2\right)$
and
$n_k \sim {\cal N}\left(0,\sigma_J^2\right)$.
Thus, the generalization error $\epsilon_g$ of the ensemble is
calculated as follows:
\begin{eqnarray}
\epsilon_{g}
&=& \int d\mbox{\boldmath $x$}dn_B \left(\prod_{k=1}^K dn_k\right)
         p(\mbox{\boldmath $x$})p(n_B)
         \left(\prod_{k=1}^K p(n_k)\right) \epsilon \\
&=& \int dv \left(\prod_{k=1}^K du_k\right) dn_B 
          \left(\prod_{k=1}^K dn_k \right) p(v,\{u_k\})
          p(n_B) \left(\prod_{k=1}^K p(n_k)\right) \nonumber \\
& & \hspace{45mm} \times \frac{1}{2}\left(v+n_B-\sum_{k=1}^K C_k 
            \left(u_k l^{m_k}+n_k \right)\right)^2 \\
&=& \frac{1}{2}\left(1-2\sum_{k=1}^K C_k l^{m_k}R^{m_k} 
       + 2\sum_{k=1}^K \sum_{k'>k}^K C_k C_{k'} l^{m_k}l^{m_{k'}}q^{m_k,m_{k'}}
     \right. \nonumber \\
& & \hspace{45mm}  \left. + \sum_{k=1}^K C_k^2(l^{m_k})^2 +\sigma_B^2
            + \sum_{k=1}^K C_k^2 \sigma_J^2 \right), \label{eqn:eg}
\end{eqnarray}
where
$v=\mbox{\boldmath $B$}\cdot\mbox{\boldmath $x$}$ and 
$u_k l^{m_k}=\mbox{\boldmath $J$}^{m_k}\cdot\mbox{\boldmath $x$}$.
We executed integration using the following:
$v$ and $u_k$ obey ${\cal N}(0,1)$.
The covariance between $v$ and $u_k$ is $R^{m_k}$,
and that between $u_k$ and $u_{k'}$ is $q^{m_k,m_{k'}}$.

\item[Nonlinear Case]
A majority vote by brothers might be 
a typical method of combining for a nonlinear model
in which the output of each student is $+1$ or $-1$.
However, to simplify the analysis
we apply the following method:
the output of the ensemble is
that of a new perceptron 
of which the connection weight
is the weighted sum 
of the normalized connection weights
$\mbox{\boldmath $J$}^{t_k}/l^{t_k}$
of brothers.
That is,
\begin{equation}
\epsilon=\Theta\left(-
\left(v-a\right)v\left(v+a\right)
\sum_{k=1}^K C_k 
u_k\right),
\label{eqn:e-n}
\end{equation}
where $C_k \ge 0$ is a weight 
of each brother $\mbox{\boldmath $J$}^{m_k}$ 
in the ensemble.
Thus, the generalization error $\epsilon_g$ of the ensemble is
expressed as follows:
\begin{equation}
\epsilon_{g}
= \int d\mbox{\boldmath $x$} 
         p(\mbox{\boldmath $x$}) \epsilon
= \int dv \left(\prod_{k=1}^K du_k\right) 
          p(v,\{u_k\})
	\Theta\left(-
	\left(v-a\right)v\left(v+a\right)
	\sum_{k=1}^K C_k u_k\right).
\label{eqn:eg-n}
\end{equation}

\end{description}

\subsection{Differential equations for order parameters,
and their solutions}
In this paper, we examine the thermodynamic limit $N\rightarrow \infty$.
To do so,
updates of Eq.(\ref{eqn:Jupdate-l}) or Eq.(\ref{eqn:Jupdate-n})
must be executed $O(N)$ times for the order 
parameters $l, R$, and $q$ to change by $O(1)$.
Thus, the continuous times $t_1,\ldots,t_K$,
which are the time steps $m_1,\ldots,m_K$ normalized
by the dimension $N$,
are introduced as the superscripts 
that represent the learning process.
To simplify the analysis, we introduced the following auxiliary 
order parameters
$r^t \equiv l^t R^t$ and $Q^{t,t'} \equiv l^t l^{t'} q^{t,t'}$.
The simultaneous differential equations
in deterministic forms \cite{NishimoriE}, 
which describe the dynamical behaviors of order parameters,
have been obtained based on the self-averaging
of thermodynamic limits as follows:
\begin{eqnarray}
\frac{dl^{t}}{dt} &=& 
\langle f^{t} u^{t}\rangle 
                   + \frac{\langle (f^{t})^2\rangle}{2l^{t}},
\label{eqn:dldt}\\
\frac{dr^{t}}{dt} &=& \langle f^{t} v^{t}\rangle, 
\label{eqn:drdt}\\
\frac{dQ^{t,t'}}{dt'} 
&=& l^t \langle f^{t'}\bar{u}^{t}\rangle,
\label{eqn:dQdt}
\end{eqnarray}
where
$t'\geq t$ and
$\bar{u}^t=\mbox{\boldmath $x$}^{t'} \cdot \mbox{\boldmath $J$}^{t}/l^{t}
\sim {\cal N}(0,1)$.
Four sample averages in Eqs.(\ref{eqn:dldt})--(\ref{eqn:dQdt})
are obtained by executing integrations
where 
$v^{t'},u^{t'}$ and $\bar{u}^{t}$
obey the triple-Gaussian distribution
$p(v^{t'},u^{t'},\bar{u}^{t})$,
for which the covariance matrix is
\begin{eqnarray}
\mbox{\boldmath $\Sigma$}
&=&
 \left(
 \begin{array}{ccc}
  1      & R^{t'}   & R^t   \\
  R^{t'} & 1        & q^{t,t'} \\
  R^t    & q^{t,t'} & 1 
 \end{array}
 \right).
\label{eqn:Sigma}
\end{eqnarray}

\begin{description}
\item[Linear Case]
The four sample averages can be easily calculated as follows:
\begin{eqnarray}
\langle f^tu^t \rangle &=& \eta (r^t/l^t-l^t), \\
\langle (f^t)^2 \rangle &=& 
\eta^2 (1+\sigma_B^2 + \sigma_J^2 +(l^t)^2 -2r^t), \\
\langle f^tv^t \rangle &=& \eta (1-r^t), \\
\langle f^{t'} \bar{u}^{t} \rangle 
&=& \eta \left(r^{t}-Q^{t,t'}\right)/l^t. 
\label{eqn:fbarut}
\end{eqnarray}

Using $R^0=0, l^0=1$ and $Q^{t,t}=(l^t)^2$ 
as initial conditions,
we can analytically solve the 
simultaneous differential equations 
Eqs.(\ref{eqn:dldt})--(\ref{eqn:dQdt})
as follows\cite{JPSJ2006c}:
\begin{eqnarray}
r^t&=& 1-e^{-\eta t}, \label{eqn:rsol} \\
(l^t)^2 &=& 
1+\frac{\eta}{2-\eta}\left(\sigma_B^2+\sigma_J^2\right)
-2e^{-\eta t}
+\left(2-\frac{\eta}{2-\eta}\left(\sigma_B^2+\sigma_J^2\right)\right)
e^{\eta(\eta-2)t}, \label{eqn:l2sol}\\
Q^{t,t'} &=& 1-e^{-\eta t}
+e^{-\eta t'}
+\left((l^t)^2-1\right)e^{-\eta (t'-t)}.
\label{eqn:Qsol}
\end{eqnarray}

Substituting Eqs.(\ref{eqn:rsol})--(\ref{eqn:Qsol}) 
into Eq.(\ref{eqn:eg}),
the generalization error $\epsilon_g$
can be analytically obtained as a function of
time $t_k,\ k=1,\ldots,K$.

\item[Nonlinear Case]
The four sample averages are obtained as follows:
\begin{eqnarray}
\langle f^{t}u^{t}\rangle \!\!\! &=& \!\!\!
   \frac{\eta}{\sqrt{2\pi}}
   \left(R^{t}\left(2\exp\left(-\frac{a^2}{2}\right)-1\right)-1\right), 
   \label{eqn:fu-n}\\
\langle (f^{t})^2\rangle \!\!\! &=& \!\!\!
  2\eta^2\left(
  \int_a^{\infty}\!
    DvH\left(\frac{R^{t}v}{\sqrt{1-\left(R^{t}\right)^2}}\right)
  \!\! + \!\!
  \int_0^a\!
    DvH\left(-\frac{R^{t}v}{\sqrt{1-\left(R^{t}\right)^2}}\right)
 \right),
 \label{eqn:f2-n}\\
\langle f^{t}v^{t}\rangle \!\!\! &=& \!\!\!
   \frac{\eta}{\sqrt{2\pi}}
   \left(2\exp\left(-\frac{a^2}{2}\right)-1-R^{t}\right), 
   \label{eqn:fv-n}\\
\langle f^{t'}\bar{u}^{t}\rangle \!\!\! &=& \!\!\!
   \frac{1}{1-\left(R^{t'}\right)^2}\left(
    \left(q^{t,t'}-R^{t'}R^t\right)\langle f^{t'}u^{t'}\rangle
    \!+\!
    \left(R^t-q^{t,t'}R^{t'}\right)\langle f^{t'}v^{t'}\rangle
   \right),
   \label{eqn:fubar-n}
\end{eqnarray}
where
$
H(u) \equiv \int_u^\infty Dx,\ 
Dx \equiv \frac{dx}{\sqrt{2\pi}}\exp\left(-\frac{x^2}{2}\right)
$.
The generalization error $\epsilon_g$
is obtained by solving Eqs.(\ref{eqn:eg-n})--(\ref{eqn:dQdt}), and
(\ref{eqn:fu-n})--(\ref{eqn:fubar-n}) numerically.

\end{description}

\section{Results and Discussion}
Figure 1 shows
examples of the dynamical behaviors of $l$ and $R$
in a linear model
obtained analytically, and the corresponding
simulation results, where $N=2,000$.
In the case of a linear model, 
many properties regarding both dynamical behaviors
and steady states can be analytically proven\cite{JPSJ2006c}.
For example,
$l$ and $\epsilon_g$ diverge unless $0<\eta<2$.
In the case of no noise,
$l$ asymptotically approaches unity
after becoming larger than unity if $0<\eta<1$
and $l$ asymptotically approaches unity
after becoming smaller than unity if $1<\eta<2$.
The larger $\eta$ is, the faster $R$ rises.
However,
the convergence of $R$ is fastest
when the learning rate satisfies $\eta=1$.
This phenomenon can be understood by the fact that
$\eta=1$ is a special condition for the gradient method
where the student uses up the information
obtained from input $\mbox{\boldmath $x$}$ \cite{JPSJ2006b}.

Figure 2 displays
some examples of the dynamical behaviors of $l$ and $R$
in a nonlinear model
obtained numerically, and the corresponding
simulation results, where $N=2,000$.
The reason why $R$ is negative,
which differs from the linear case,
is that the threshold $a$ of a nonmonotonic teacher
is greater than the critical value $a_C=\sqrt{2\ln 2}\simeq 1.18$
\cite{Inoue}.
This is not essential in this paper.
When the learning rate $\eta$ is relatively large,
the dynamical behavior of $R$ is monotonic;
however, when $\eta$ is small,
the dynamical behavior of $R$ becomes nonmonotonic.
That is, $|R|$ asymptotically approaches a steady value
after reaching its maximal one.
The steady value is not dependent upon $\eta$.

\begin{figure}[htbp]
\begin{center}
\includegraphics[width=0.60\linewidth,keepaspectratio]{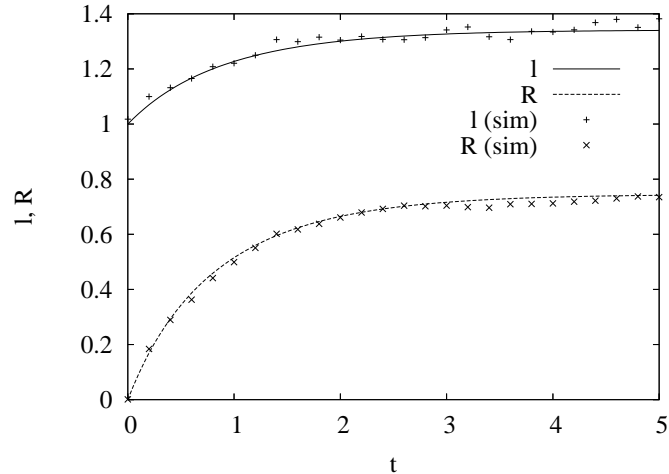}\\
\caption{Dynamical behaviors of $l$ and $R$ in the linear case.
$\eta=1.0,\ \sigma_B^2=0.3,\ \sigma_J^2=0.5$,}
\end{center}
\label{fig:lR2}
\end{figure}

\begin{figure}[htbp]
\begin{center}
\includegraphics[width=0.60\linewidth,keepaspectratio]{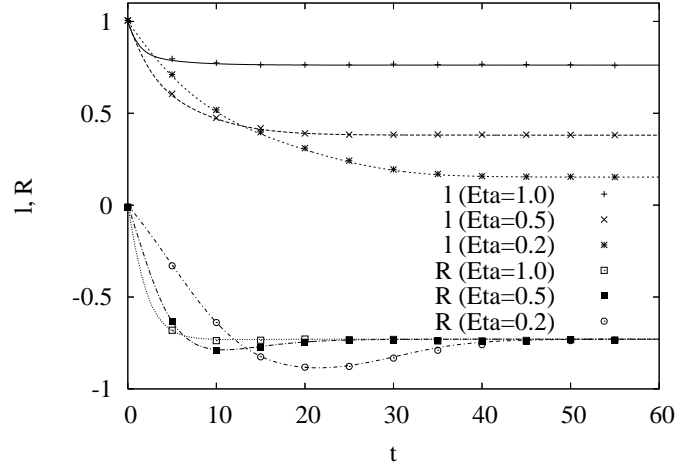}\\
\caption{Dynamical behaviors of $l$ and $R$ in the nonlinear case. $a=2.0$.}
\end{center}
\label{fig:lRa20}
\end{figure}

\begin{figure}[htbp]
\begin{minipage}{.500\linewidth}
\begin{center}
\includegraphics[width=1.000\linewidth,keepaspectratio]{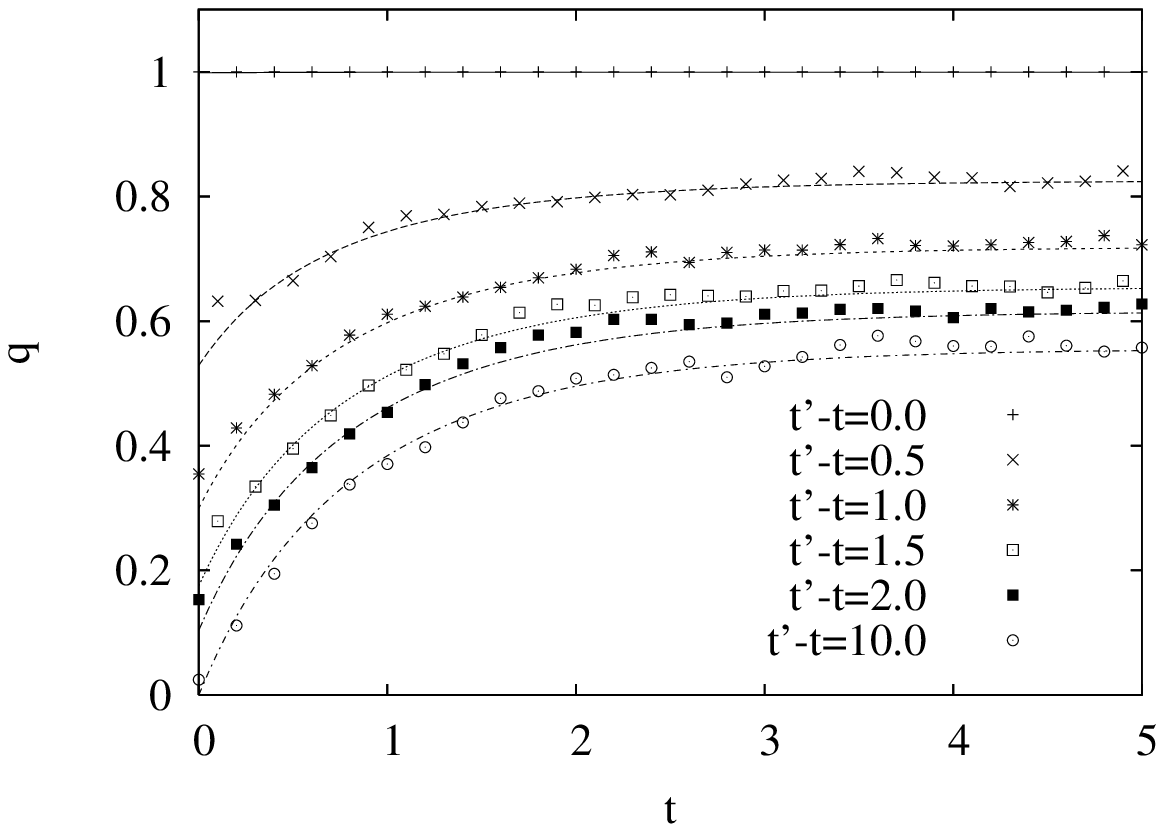}\\
\end{center}
\end{minipage}
\begin{minipage}{.500\linewidth}
\begin{center}
\includegraphics[width=1.000\linewidth,keepaspectratio]{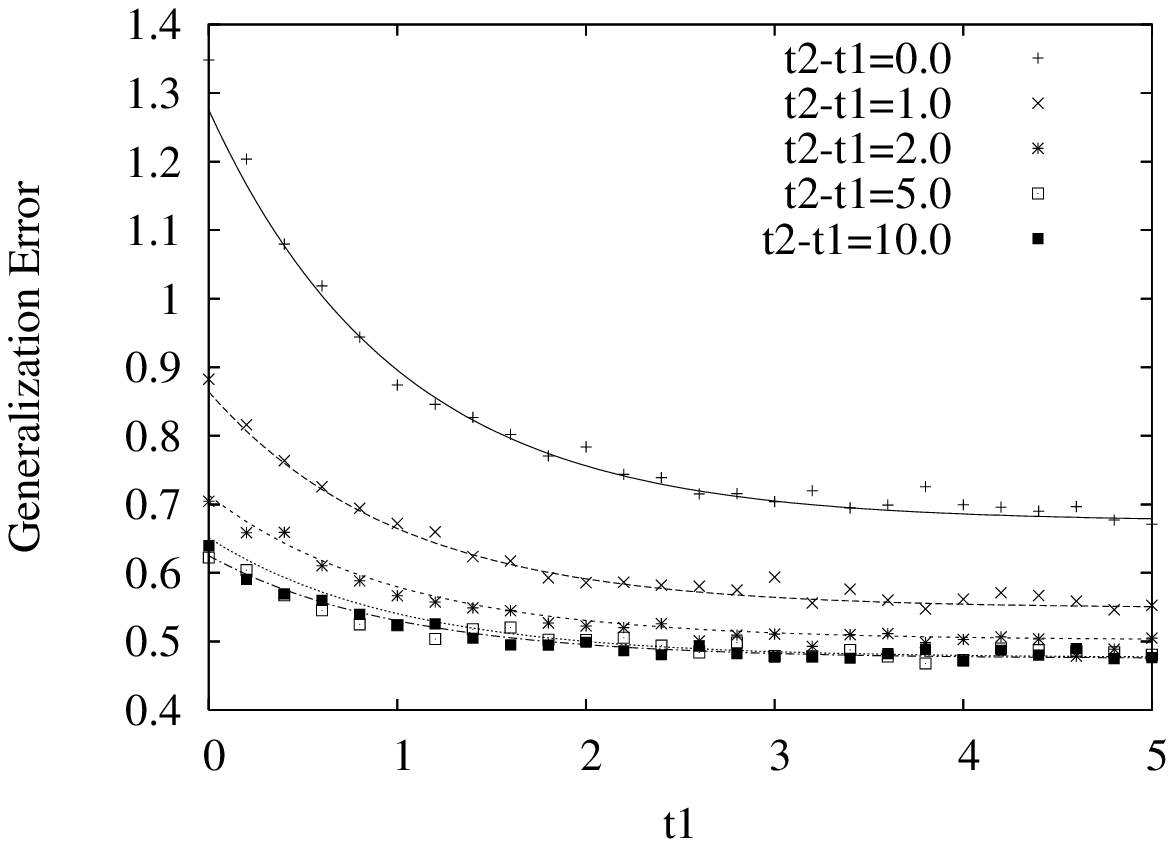}
\end{center}
\end{minipage}
\caption{Dynamical behaviors of $q$ and $\epsilon_g$ in 
\underline{a linear case}.
$\eta=1.0,\ \sigma_B^2=0.3,\ \sigma_J^2=0.5,\ K=2,\ C_k=1/K$.}
\label{fig:qegl}
\end{figure}

Figure \ref{fig:qegl} (left)
presents
some examples of the dynamical behaviors of $q$
of a linear model
obtained analytically, and the corresponding
simulation results, where $N=2,000$.
For a linear model, 
$q$ increases monotonically when $t$ increases,
and 
increases monotonically when $t'-t$ decreases.
Furthermore, $q^{t,t'}$ converges to a smaller value 
than unity in the case of $t'-t\neq 0.0$.
This means the student itself continues to move
after the order parameters 
reach a steady state.
Figure \ref{fig:qegl} (right)
shows the relationship
between $t_1$ and $\epsilon_g$ in the case of 
constant $t_2-t_1$ and $K=2$.
Here, 
$\epsilon_g$ increases monotonically, remains constant, or
decreases monotonically depending on the values of $\eta$.

The behaviors of $\epsilon_g$ when 
the leading time $t_1 \rightarrow \infty$
and 
the time interval $t_{k+1}-t_k \rightarrow \infty$
can be theoretically obtained in the case of a linear model
as follows \cite{JPSJ2006c}:
$\epsilon_g$ decreases
as $\eta$ decreases regardless of $K$.
When the weights are uniform or $C_k=C=1/K$
and $K=\infty$,
$
\mbox{\boldmath $B$}=\lim_{K\rightarrow \infty}
\frac{1}{K}\sum_{k=1}^K \mbox{\boldmath $J$}^{t_k}
$.
This means the generalization error equals
the residual error caused by teacher's noise
$n_B$.
On the other hand,
the generalization error $\epsilon_g$ of $K=\infty$
is $\frac{1}{4}$ times of that of $K=1$ when 
the learning rate satisfies $\eta=1$, 
the uniform weights $C_k=1/K$,
$\sigma_B^2=\sigma_J^2$, 
$t_1 \rightarrow \infty$, and $t_{k'}-t_k \rightarrow \infty$.
Since the generalizaion error $\epsilon_g$ of 
the conventional 
space-domain ensemble learning with 
$K=\infty$, $\eta=1$, $C_k=1/K$ and 
$\sigma_B^2=\sigma_J^2$
is $\frac{1}{2}$ times of that of $K=1$ \cite{Hara}, 
we can say the time-domain
ensemble learning is twice as effective as 
the conventional space-domain ensemble learning.
This difference can be explained as follows:
In conventional space-domain
ensemble learning, the similarities among students
become high since all students use the same examples for learning.
In time-domain ensemble learning,
on the other hand,
the similarities among brothers become low
since all brothers use almost totally different examples for learning.

\begin{figure}[htbp]
\begin{minipage}{.500\linewidth}
\begin{center}
\includegraphics[width=1.000\linewidth,keepaspectratio]{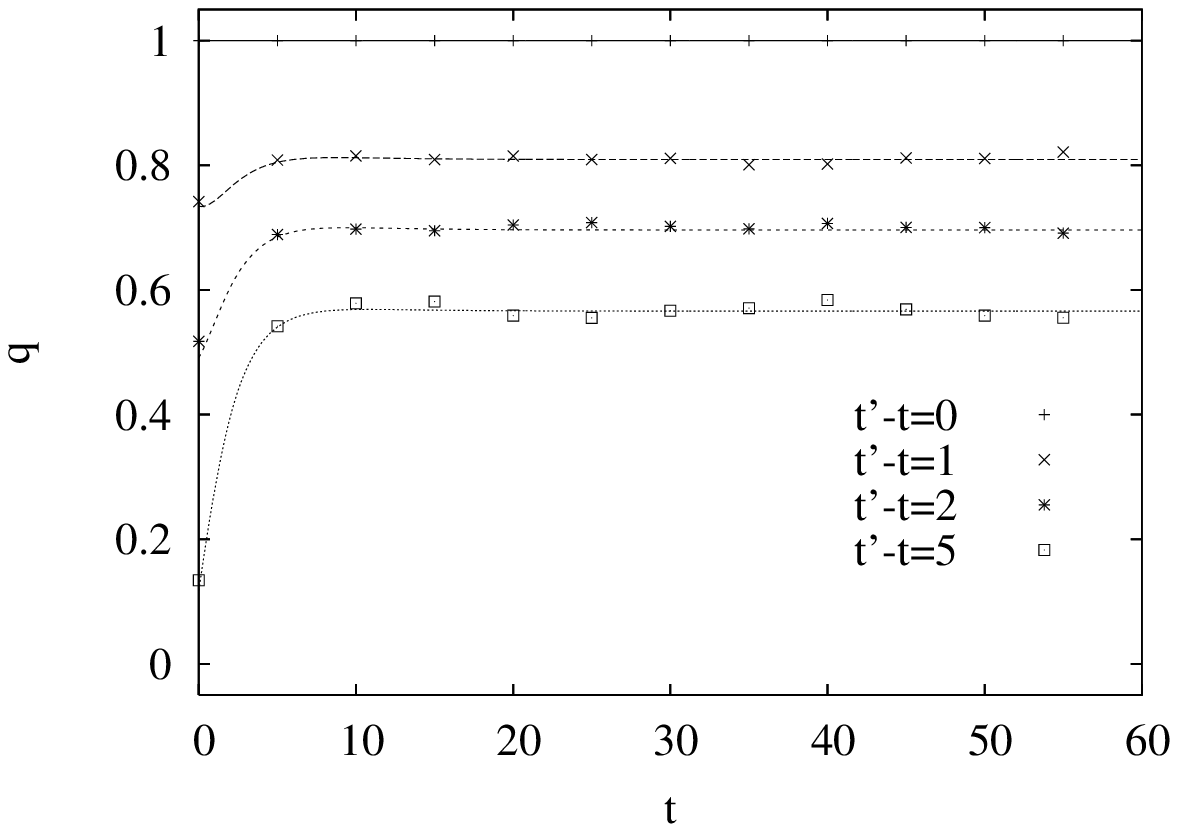}\\
\end{center}
\end{minipage}
\begin{minipage}{.500\linewidth}
\begin{center}
\includegraphics[width=1.000\linewidth,keepaspectratio]{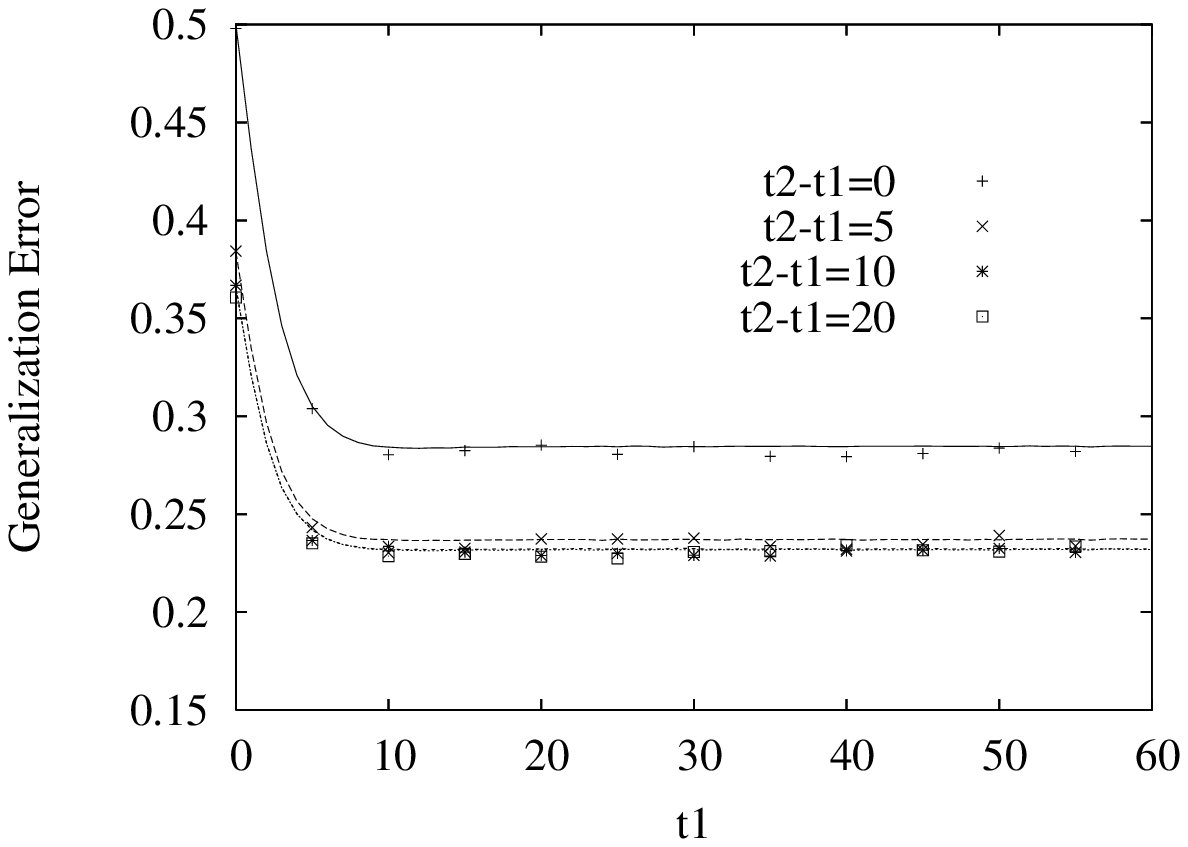}
\end{center}
\end{minipage}
\caption{Dynamical behaviors of $q$ and $\epsilon_g$ in 
\underline{a nonlinear case}.
\underline{$\eta=1.0$},$\ a=2.0,\ K=2,\ C_k=1/K$.}
\label{fig:qegn10}
\end{figure}

\begin{figure}[htbp]
\begin{minipage}{.500\linewidth}
\begin{center}
\includegraphics[width=1.000\linewidth,keepaspectratio]{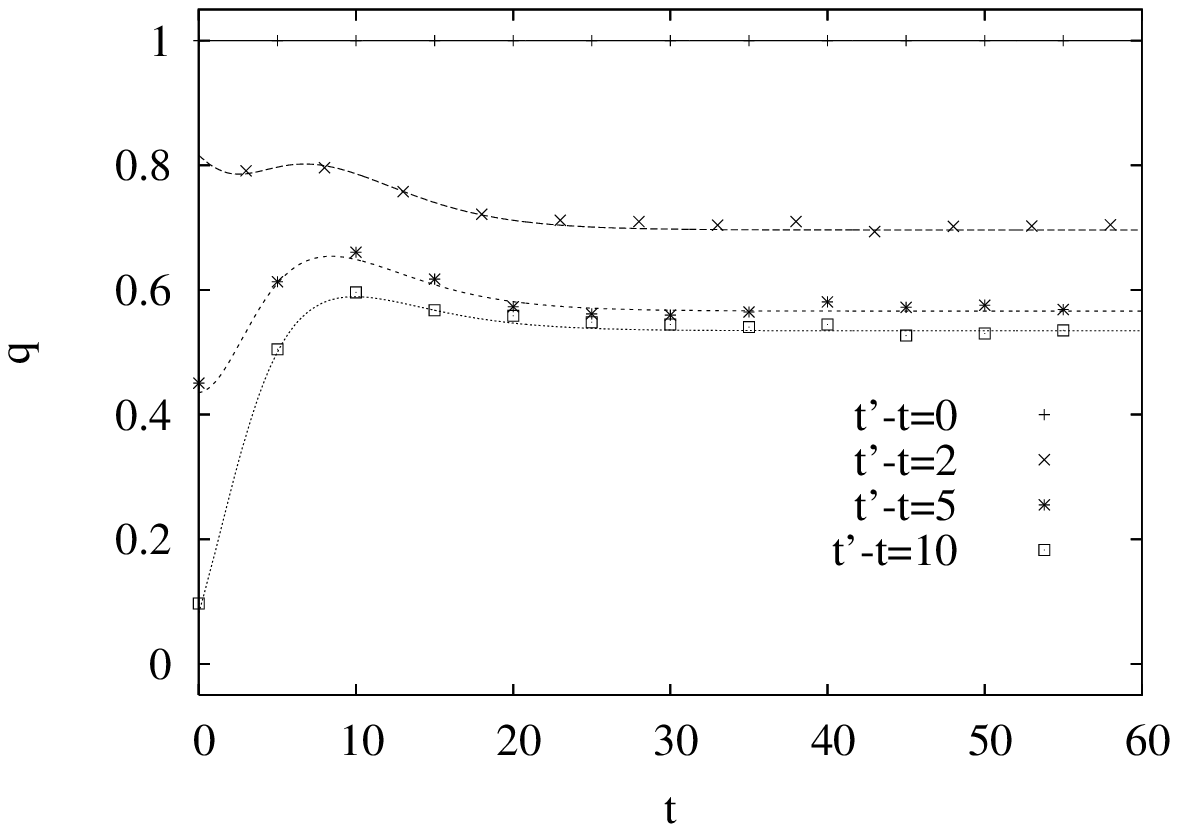}\\
\end{center}
\end{minipage}
\begin{minipage}{.500\linewidth}
\begin{center}
\includegraphics[width=1.000\linewidth,keepaspectratio]{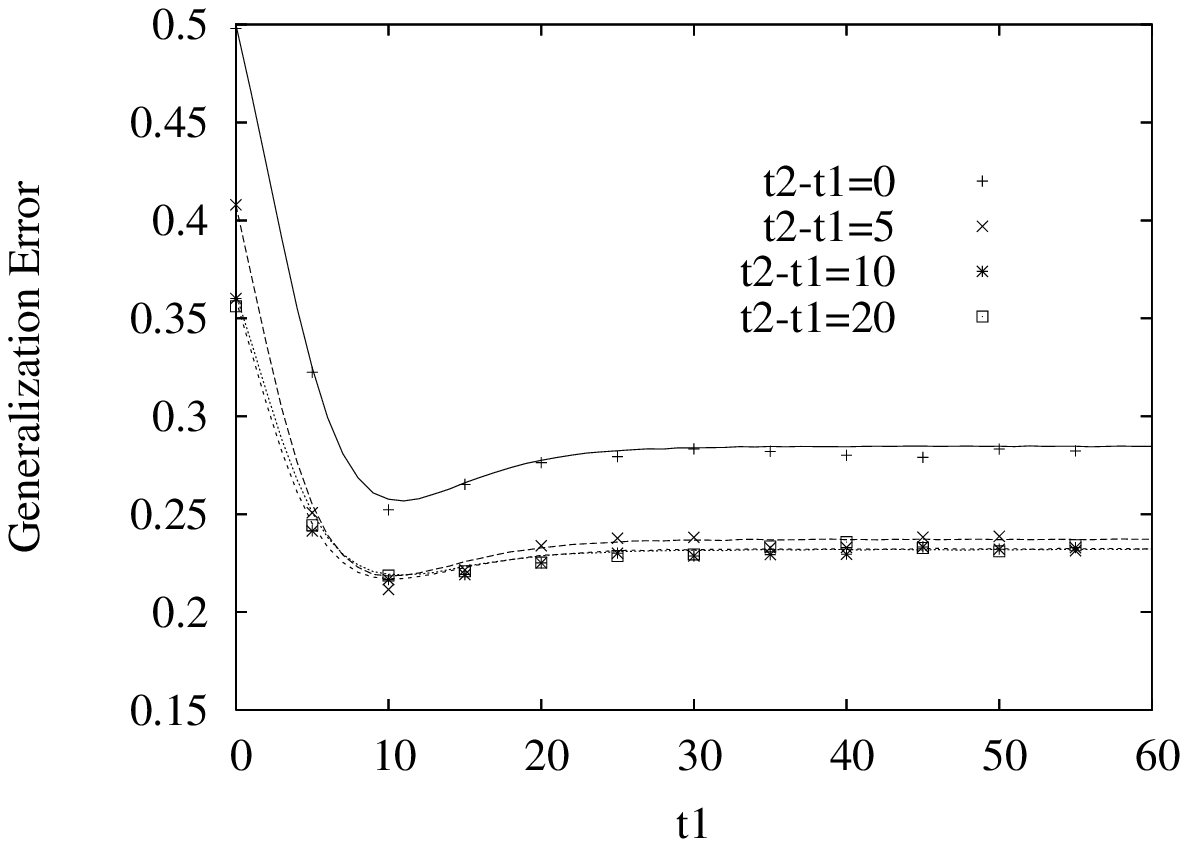}
\end{center}
\end{minipage}
\caption{Dynamical behaviors of $q$ and $\epsilon_g$ in 
\underline{a nonlinear case}.
\underline{$\eta=0.5$},$\ a=2.0,\ K=2,\ C_k=1/K$.}
\label{fig:qegn05}
\end{figure}

\begin{figure}[htbp]
\begin{minipage}{.500\linewidth}
\begin{center}
\includegraphics[width=1.000\linewidth,keepaspectratio]{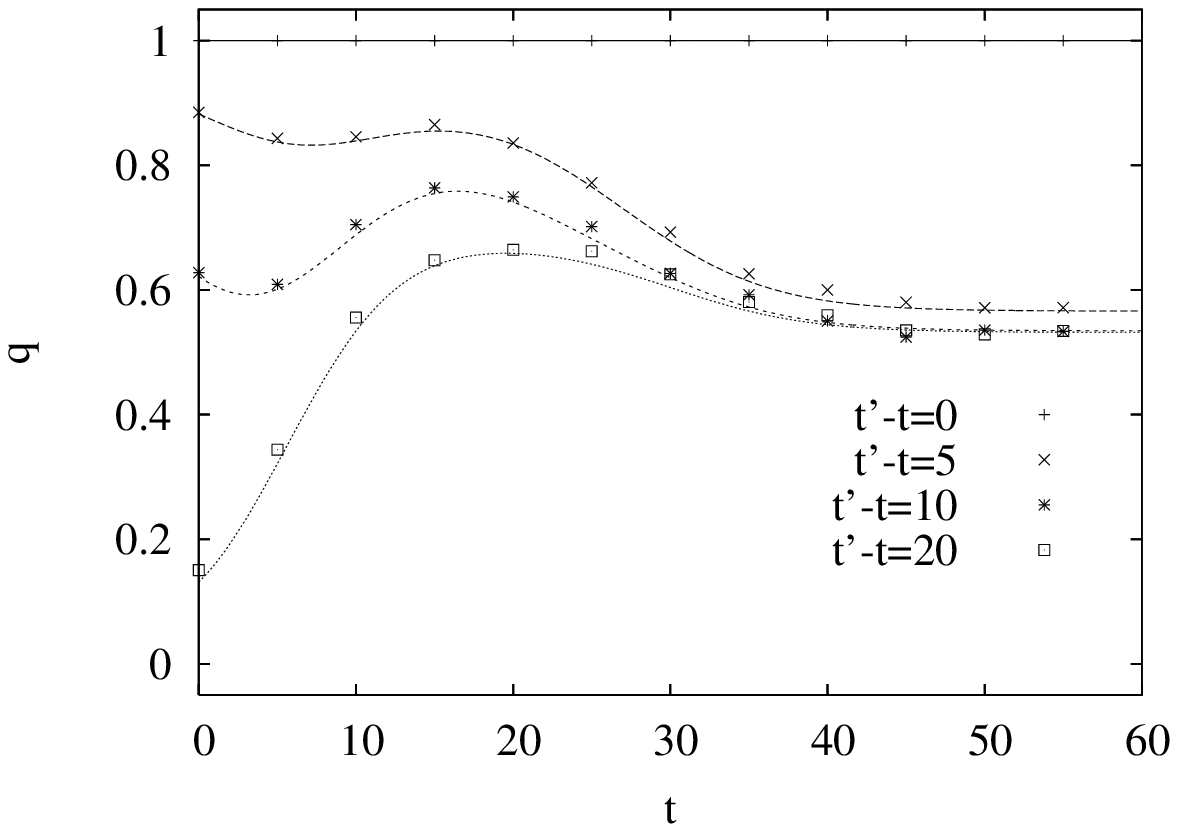}\\
\end{center}
\end{minipage}
\begin{minipage}{.500\linewidth}
\begin{center}
\includegraphics[width=1.000\linewidth,keepaspectratio]{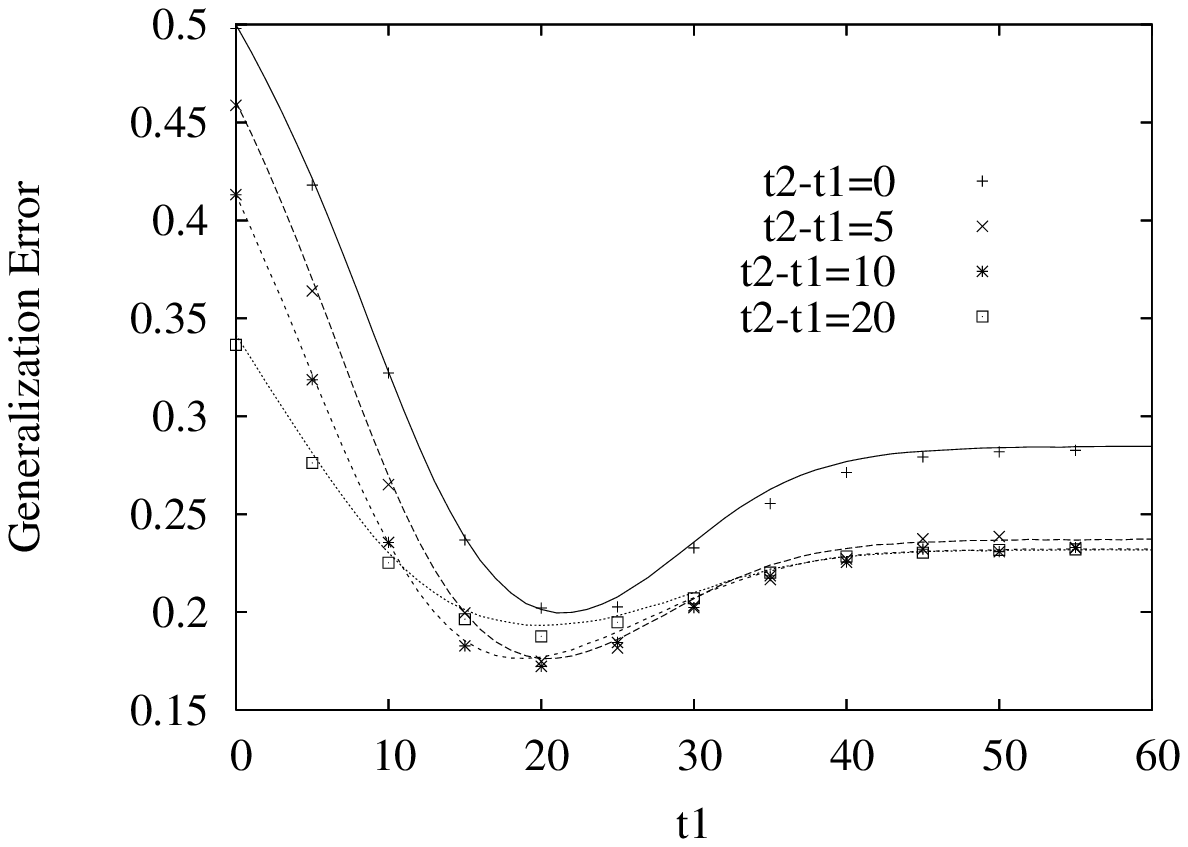}
\end{center}
\end{minipage}
\caption{Dynamical behaviors of $q$ and $\epsilon_g$ in 
\underline{a nonlinear case}.
\underline{$\eta=0.2$},$\ a=2.0,\ K=2,\ C_k=1/K$.}
\label{fig:qegn02}
\end{figure}

Figures \ref{fig:qegn10}--\ref{fig:qegn02} (left)
show
some examples of the dynamical behaviors of $q$
for a nonlinear model
obtained numerically, and the corresponding
simulation results, where $N=2,000$.
These figures indicate that
$q$ for a nonlinear model 
behaves nonmonotonically for $t$
when $\eta$ is small.
Figures \ref{fig:qegn10}--\ref{fig:qegn02} (right)
show the relationship between $t_1$ and $\epsilon_g$
in the case of constant $t_2-t_1$ and $K=2$.
The steady value of $\epsilon_g$
is dependent upon $t_2-t_1$
and is not dependent upon $\eta$.
However, when $\eta$ is small,
$\epsilon_g$ behaves nonmonotonically for $t_1$
and has the minimal value
shown in 
Figs.\ref{fig:qegn05} (right) and \ref{fig:qegn02} (right).
This phenomenon can be considered a kind of 
over-learning.
Figure \ref{fig:qegn02} (right) shows
that the minimal value of $\epsilon_g$
decreases when $t_2-t_1$ increases as 
$0\rightarrow 5\rightarrow 10$ and
increases when $t_2-t_1$ increases as 
$10\rightarrow 20$.
This means that $t_2-t_1$ has an optimum value.
Figures \ref{fig:qegn05} (right) 
and \ref{fig:qegn02} (right) reveal that
the smaller the learning rate $\eta$ is,
the smaller the minimal value of $\epsilon_g$ is.
However, if $\eta$ is too small,
the learning becomes slow.
Therefore,
if the aim is to decrease the generalization error
$\epsilon_g$,
we should use the smallest $\eta$ that is
possible from the viewpoint of learning speed,
set $t'-t$ to the optimum value,
and
stop the learning at an adequate time step.

\section{Conclusion}
We have analyzed
the generalization performances regarding 
time-domain ensemble learning of 
both a linear model and a nonlinear model.
Analyzing 
within the framework of online learning 
using a statistical mechanical method,
we have demonstrated the qualitatively different 
behaviors between the two models.
In a linear model,
the dynamical behaviors of the generalization error
are monotonic.
We have analytically shown that time-domain
ensemble learning is twice as effective as 
conventional ensemble learning.
Furthermore, the generalization error of 
a nonlinear model
exhibits nonmonotonic dynamical behaviors
when the learning rate is small.
We have numerically shown that the generalization performance 
can be remarkably improved by
using this phenomenon together with the divergence of 
students in the time domain.

\subsubsection*{Acknowledgments}
This research was partially supported by the Ministry of Education, 
Culture, Sports, Science, and Technology of Japan, 
with Grants-in-Aid for Scientific Research
14084212, 15500151, 16500093, and 18500183.

\small{

\end{document}